\documentclass[manuscript]{aastex}

\slugcomment{Not to appear in Nonlearned J., 45.}

\shorttitle{Quasi-periodicities of PKS 2155-304}
\shortauthors{A. Sandrinelli, S. Covino, A. Treves}

\begin{document}


\title{Quasi-periodicities of  the BL Lac Object PKS 2155-304}


\author{A. Sandrinelli}
\affil{\textit{Universit\`a degli Studi del$\l'$Insubria, Via Valleggio 11, I-22100 Como, Italy\\
INAF - Osservatorio Astronomico di Brera, Via Emilio Bianchi 46, I-23807 Merate, Italy.\\
INFN - Istituto Nazionale di Fisica Nucleare, Sezione Milano Bicocca, \\
Dipartimento di Fisica G. Occhialini, Piazza della Scienza 3, I-20126 Milano, Italy }}
\email{asandrinelli@yahoo.it}

\author{S. Covino}
\affil{\textit{
INAF - Osservatorio Astronomico di Brera, Via Emilio Bianchi 46, I-23807 Merate, Italy}}

\author{A. Treves}
\affil{\textit{Universit\`a degli Studi del$\l'$Insubria, Via Valleggio 11, I-22100 Como, Italy\\
INAF - Osservatorio Astronomico di Brera, Via Emilio Bianchi 46, I-23807 Merate, Italy\\
INFN - Istituto Nazionale di Fisica Nucleare, Sezione Milano Bicocca, \\
Dipartimento di Fisica G. Occhialini, Piazza della Scienza 3, I-20126 Milano, Italy }}

\begin{abstract}

We have searched for periodicities  in our VRIJHK photometry of PKS 2155-304, which covers the years 2005-2012.
 A peak of the Fourier spectrum with high significance is found at T$\sim$315 days, confirming the recent 
  findings by \cite{Zhang2014}. 
 The examination of the gamma-ray light curves from the\textit{ Fermi} archives yields a significant signal at $\sim$ 2T, 
which, while nominally significant, involves data spanning only  $\sim$ 6T.
 Assuming a black hole mass of $10^{9}\  M_{\odot}$ the Keplerian distance corresponding to the quasi-period T is $\sim10^{16}$ cm,
  about 50 Schwarzschild radii.

\end{abstract}


\section{Introduction}

Variability is one of the main tools for constraining models of BL Lac objects.  The time scales can be so short 
(e.g. hours, minutes) that, even taking into account relativistic beaming effects, they can shed light on sizes
 comparable with the gravitational radius of the black hole that is supposed to lie at their center  \citep[e.g.][]{Urry1991}. 
 Variability on month/year time scale is a probe of the relativistic jet structure indicating for instance
  a global helicoidal pattern,  and a possible stratification in the direction perpendicular to the jet axis
  \citep[e.g.][]{Marscher2008}.
     This kind of variability is essentially non periodic, describing turbulent mechanisms that may 
  ultimately have been triggered by some instability in the accretion process \citep[e.g.][]{Ulrich1997}.

  Up to now the only strong claim of periodicity in emission of a BL Lac object was raised in the
    case of OJ 287 where, on the basis of optical records extending for a century, a period of 12 
    years was suggested \citep{Sillampaa1988}. It may be interpreted as a Keplerian period of a system of two 
    black holes of masses 1.7 x $10^{10} \ M_{\odot}$ and 1.4 x $10^{8} \ M_{\odot}$ and  
 orbital semi-major axis of  0.056 pc   \citep{Lehto1996}. Note, however, that this
    periodicity is  still widely discussed    \citep[e.g.][]{Hudec2013}.

In this letter we concentrate on the case of the BL Lac object PKS 2155-304 (z=0.116, V$\sim$ 13), 
which is a prototype of the class, and one of the most intensely observed objects, since its discovery
 in the seventies. 
 It is a highly variable source, and numerous multifrequency observations from radio to TeV 
 are now available \citep[e.g.][and references therein]{Aharonian2007,Foschini2008}.   
Possible quasi-periodicities  on various time scales have been reported.
  \cite{Urry1993} found a repetition  time scale  of 0.7 days in ultraviolet and optical bands 
  in a  campaign lasting about one month. 
  \cite{Lachowicz2009} using few days of data from  XMM-\textit{Newton} proposed a quasi-periodicity of 4.6 h, 
 with a rather convincing sinusoidal folded light curve. 
\cite{Fan2000} collected a miscellaneous set of data from  $\sim$ 17 years of UBVRI observations,
 and found evidence of quasi-periodicity at 4 and 7 years.  
 
   More recently \cite{Zhang2014} made a collection of rather  inhomogeneous optical data over the
    last 35 years. In particular they produced an R-band light curve  with $\sim$ 8000 points deriving from
   the photometrical data of 25 different astronomical groups.  
   This light curve was searched for periodicity with various procedures, namely, epoch-folding
    \citep{Leahy1983}, the Jurkevich method \citep{Jurkevich1971} and the discrete
   correlation function \citep{Edelson1988},  that yielded, with high significance, a  quasi-period T $\sim$ 317 days.
   
In Section 2 we consider  for periodicities our homogeneous optical-NIR photometry of PKS 2155-304
 \citep{Sandrinelli2014}, which was not included in  \cite{Zhang2014} data collection.
We then examine the archived \textit{Fermi}  $\gamma$-ray curves (Section 3).  Conclusions are drawn in Section 4.

\section{REM data}
 
 Our photometric study  of PKS 2155-304 was obtained with  the \textit{Rapid Eye Mounting Telescope}
 \citep[REM, 60 cm, ][]{Zerbi2004,Covino2004} in the VRIJHK bands from 2005/05/18 (MJD 53508) to 
 2012/05/29 (MJD 56076).  Results and details about reduction and analysis procedures  are thoroughly discussed 
 in \cite{Sandrinelli2014}. 
Instrumental  magnitudes were obtained via aperture photometry.
 Calibration was performed by means of  
comparison stars in the field reported in
 \textit{Two Micron All Sky Survey 
Catalog}\footnote{\texttt{http://www.ipac.caltech.edu/2mass/}} 
\citep[2MASS,][]{Skrutskie2006} for NIR images and in \cite{Hamuy1989} for the optical.
Check stars were used in all CCD frames. Though the photometry is fully automatized,
a direct eye check of the data was performed. Several frames were excluded, because the target or 
the reference stars were at the border, or other obvious problems were apparent.

\begin{figure*}
\centering
   \includegraphics[width=.41\textwidth]{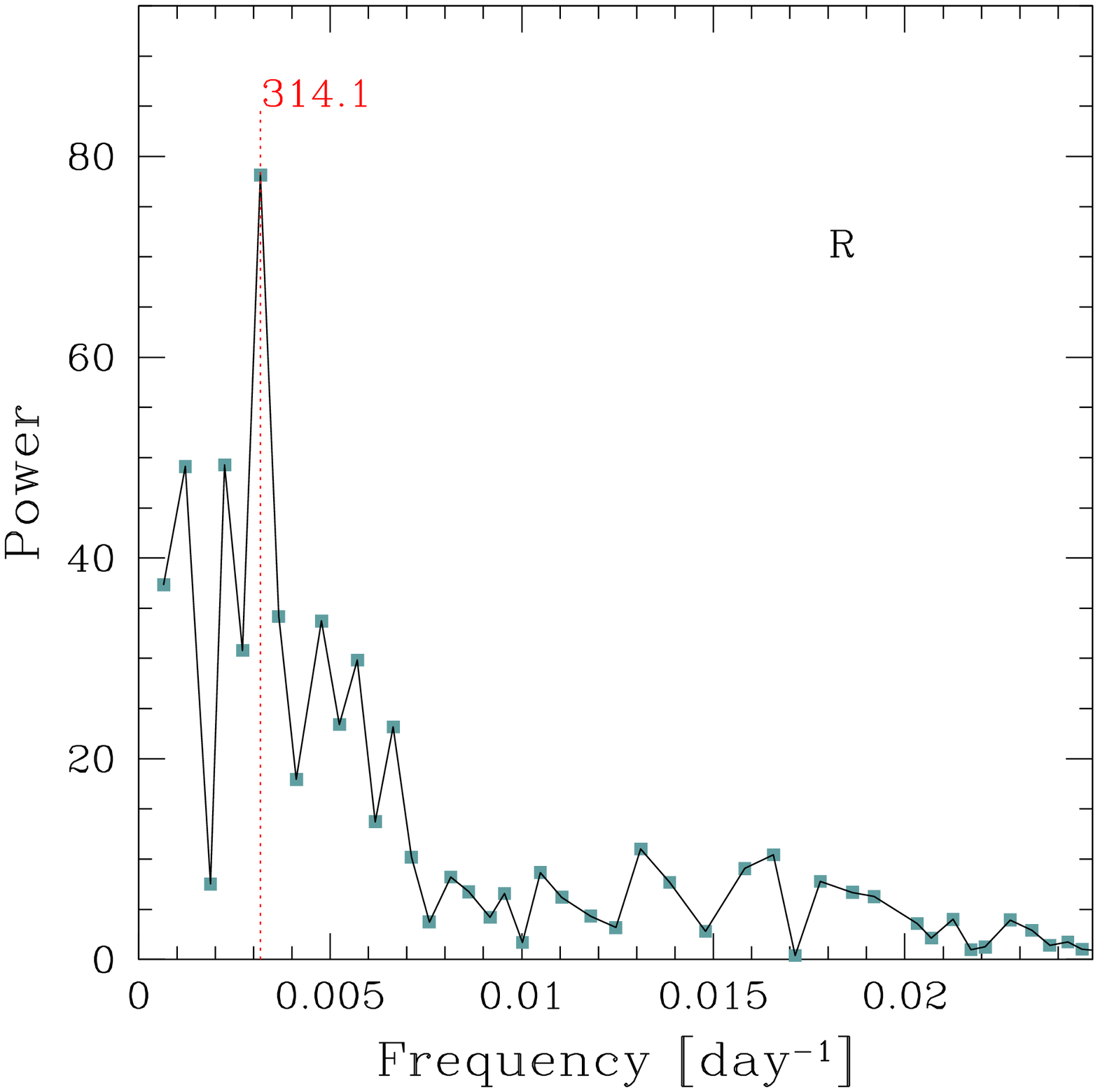}
     \includegraphics[width=.41\textwidth]{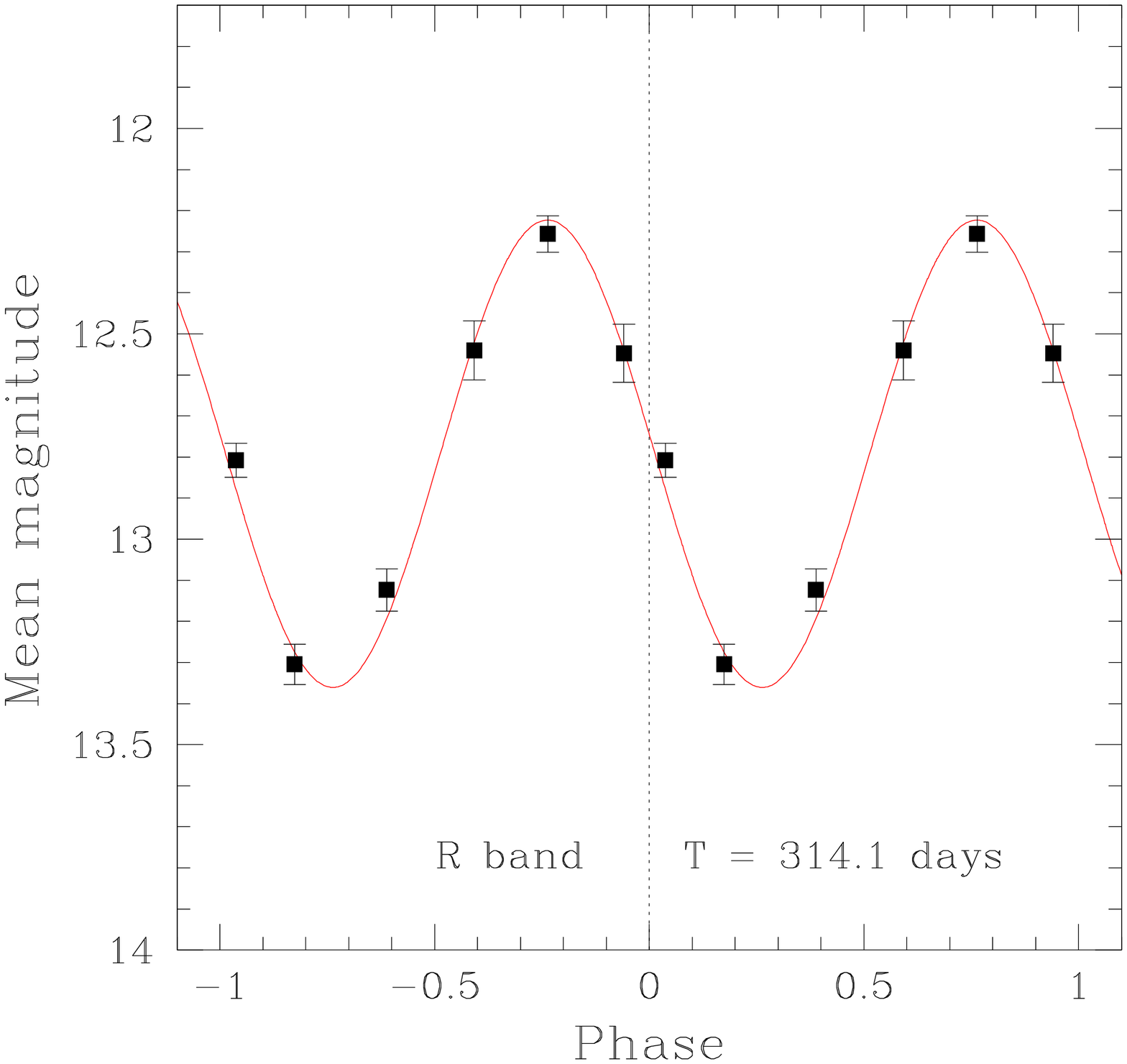}
   \includegraphics[width=.41\textwidth]{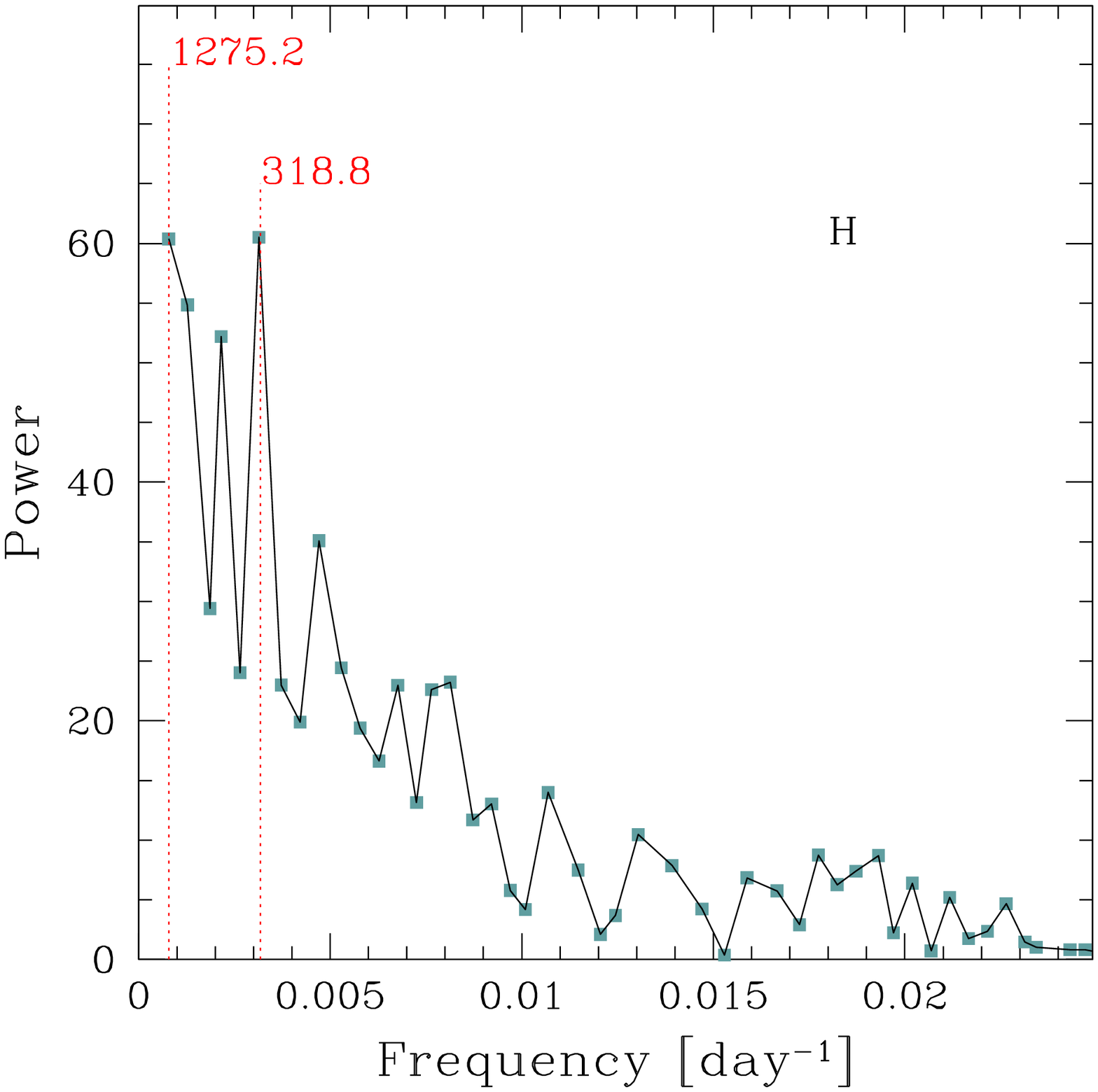}
   \includegraphics[width=.41\textwidth]{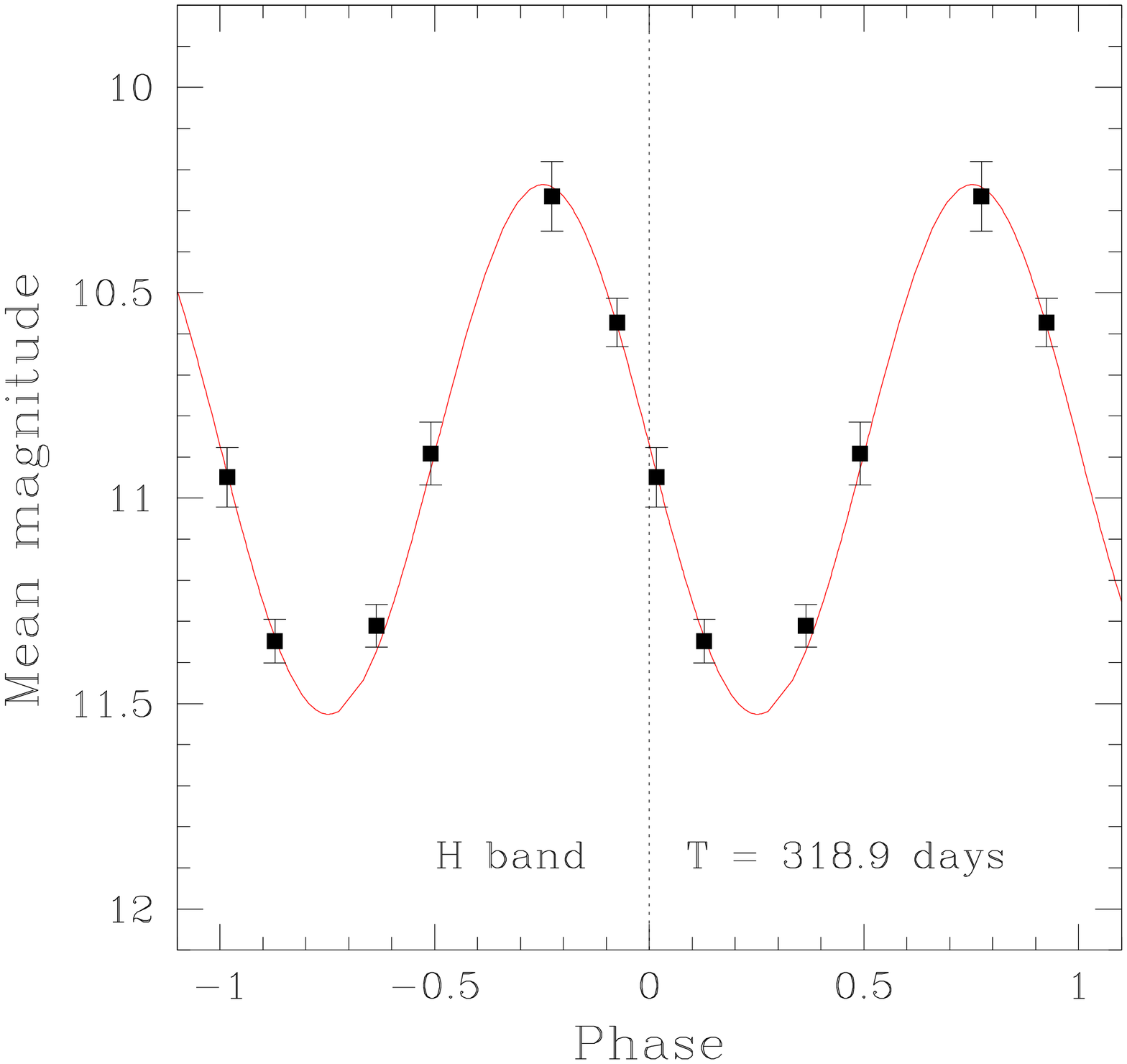}
  \includegraphics[width=.41\textwidth]{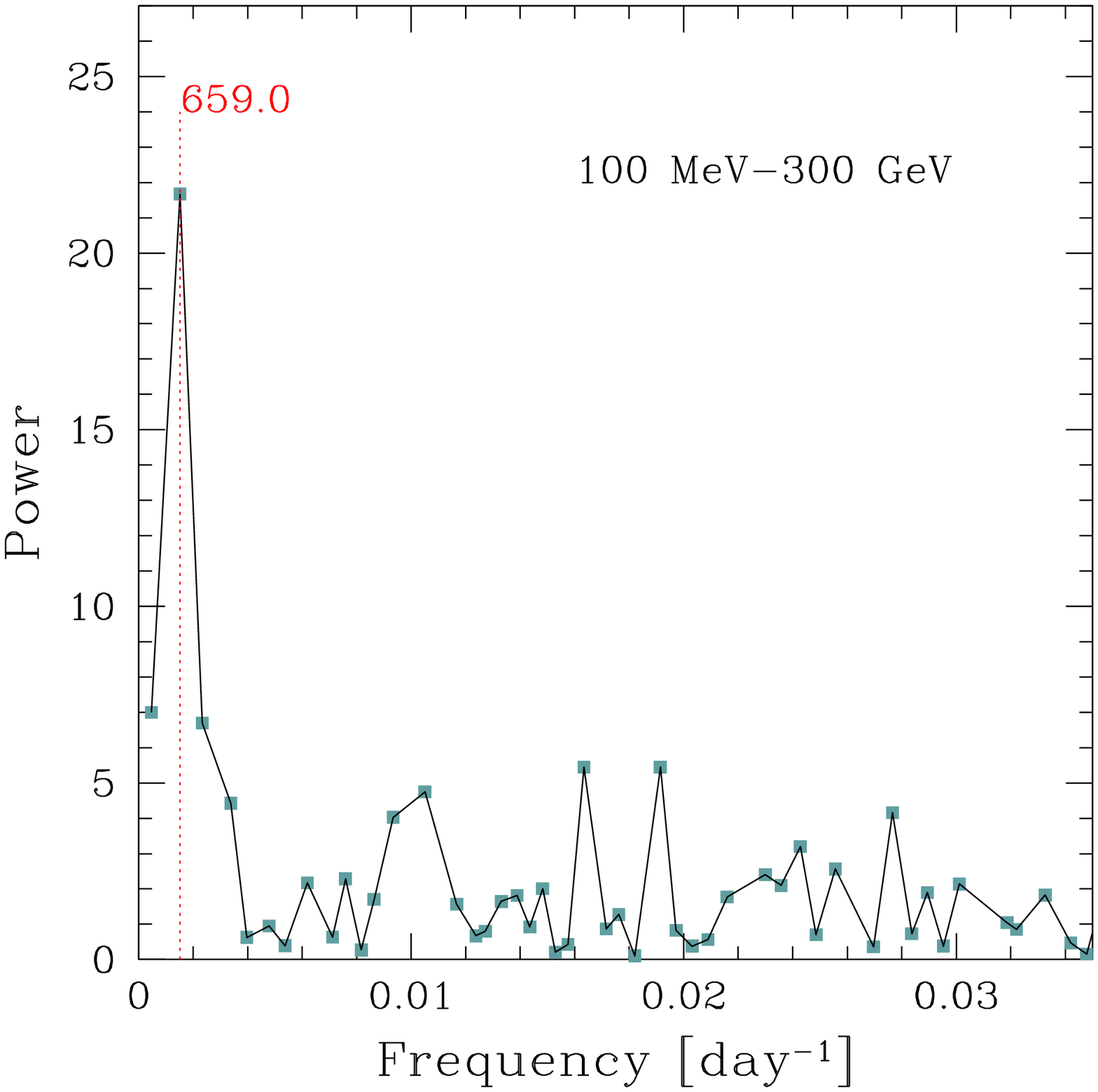}  
  \includegraphics[width=.41\textwidth]{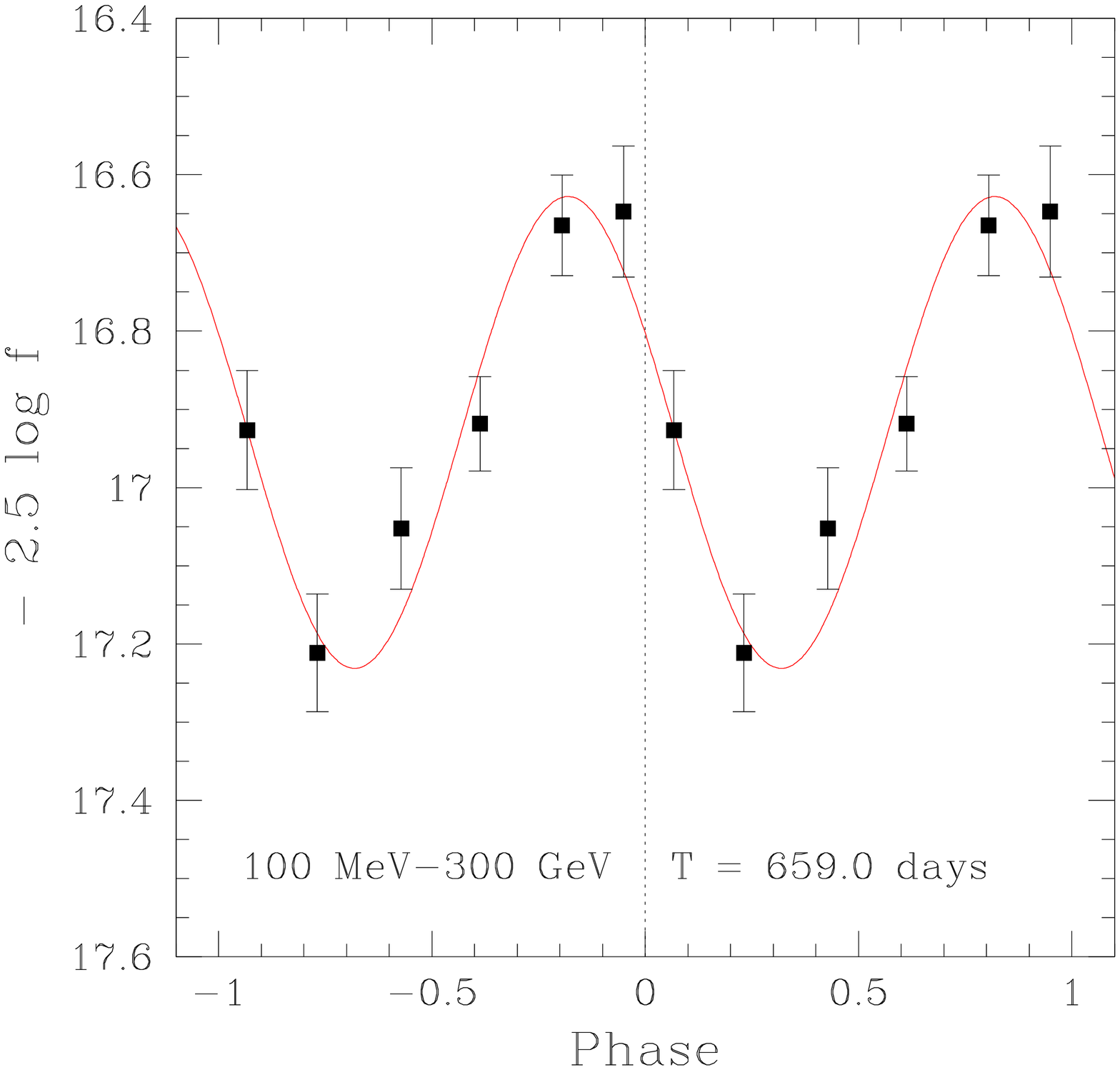}  
 \caption{ \label{pow1}  
   \footnotesize
              \textit{Left panels:} Power spectra, computed with the DC DFT approach, in R and H bands and in  
             100 MeV - 300 GeV range.  Periods in days corresponding  to the prominent peaks are 
              marked.
         \textit{Right panels:} Folded light curves. 
          Error bars are the standard errors of the average. Flux f is in photon $\textrm{cm}^{-2}\  \textrm{s}^{-1}$.
                     The initial time for folding cycles is the same for all the curves  (2009/07/28, 55040 MJD).
}
            \end{figure*}

  Since we are interested    in a long
 periodicity, we considered light curves based on nightly averages.
We searched for periodicity using the Date Compensated Discrete Fourier Transform 
 \citep[DC DFT,][]{Ferraz1981,Foster1995}   and the Lomb-Scargle (LS) algorithms \citep{Scargle1982}. 
 Results are reported in Table \ref{period} and in Figure \ref{pow1}. 
The Fourier spectra of the photometry in the VIJK filters  are very similar to those in R and H,
 and are not reported in the Figure.

   Our most prominent peaks are in the $T_{1}$ = 306-319 days range and their significance is high. 
   In the NIR Fourier spectra other noticeable peaks appear at $\sim$ 461 days and at $\sim$ 1270 days
    ($\sim 4 \cdot T_{1}$).
  Folded light curves based on DC-DFT\footnote{\texttt{http://www.aavso.org}} periods 
   are shown in Figure \ref{pow1}, where data are divided  in   $T_{1}/6$ width bins.
 The folded light curves  are clearly sinusoidal, with an amplitude of  $\sim$ 0.6 mag.  
Therefore we basically confirm the results of \cite{Zhang2014} 
 with completely independent optical and NIR data.

\begin{table*}
\begin{center}
\caption{Period analysis.}
\label{period}
\small
\begin{tabular}{@{}ccc|ccc|ccccccc}
\hline\hline
			&			&			&\multicolumn{3}{c|}{LS}&\multicolumn{7}{c}{DC-DFT}	\\
\hline
Band		&$N_{obs}$ 	&$N_{night}$	&T	       	&Power	&p-value		&T	       		&Power	&A				 	&p-value	\\
			&	  		&     			&[days]	&		&			&[days]		&		&[mag]				&		\\
(a)			&(b)	  		&(c)	     		&(d)		&(e)		&(f)			&(g)			&(h)		&(i)					&(j)		\\
 \hline													
&&&&&&&&&&&&\\
V	       		&1938	&275			& 306	&75.42	&$< 5\cdot 10^{-7}$	&309		&81.15	&0.65				&$< 1\cdot 10^{-6}$ \\
R 			&1903	&297			& 309	& 74.54	&$< 5\cdot 10^{-7}$	&314		&78.14	&0.60 				&$< 1\cdot 10^{-6}$ \\
I			&1743	&281			& 315	& 77.85	&$< 5\cdot 10^{-7}$	&316		&76.24	&0.58 				&$< 1\cdot 10^{-6}$ \\
J			&547	&274			&317 	& 62.21	&$< 5\cdot 10^{-7}$	&316		&62.99	&0.65 				&$< 1\cdot 10^{-6}$ \\
H			&646	&258			& 314 	& 62.32	&$< 5\cdot 10^{-7}$	&319		&60.50	&0.64 			 	&$< 1\cdot 10^{-6}$ \\
K			&455	&243			&318	& 61.09	&$< 5\cdot 10^{-7}$	&317		&62.03	&0.71				&$< 1\cdot 10^{-6}$ \\
&&&&&&&&&&&&\\
\hline
&&&&&&&&&&&&\\
100 MeV - 300 GeV & --	&305			&650	&22.02	&$8.4\cdot 10^{-7}$		&659		&21.67	&0.30 			&$1.6\cdot 10^{-5}$ \\
 300 MeV -1 GeV    & --	&295			&625 	&12.28	&$1.4\cdot 10^{-2}$		&621 		&12.26	&0.16 			&$5.7\cdot 10^{-5}$ \\	
&&&&&&&&&&&&\\
\hline
\end{tabular}
\end{center}
{\small
Notes:
(a) Band or range of energy.
(b) Number of observations.
(c) Number of night-average photometric points, or one-week integrations for $\gamma$-rays.
  (d) Period corresponding to the maximum frequency in LS periodogram.
   (e) Power of maximum frequency in LS periodogram.
  (f)  LS p-value.
  (g) Period corresponding to the maximum frequency in DC-DFT technique.
   (h) Power of maximum frequency in  DC-DFT technique.
 (i) Amplitude.
(j) DC-DFT p-value.
 }
 \end{table*}

  \section{Fermi data}

PKS 2155-304 is a well known $\gamma$-ray emitter \citep[$\textrm{average photon flux f}_{1-100 \ GeV}= 2.35  \pm 0.06  \cdot 10^{-8}$ photon cm$^{-2}\  \textrm{s}^{-1}$   
and  photon index  $\Gamma_{100\  MeV-100 \ GeV} = 1.84 \pm 0.02$,][]{Nolan2012}.
We considered the \textit{Fermi}\footnote{\texttt{http://fermi.gsfc.nasa.gov/ssc/data/access/lat/msl\_lc/}} light 
curves 
as provided by the automatic standard analysis procedure.
Light curves cover the interval 
2008/08/06 (54684 MJD) to  2014/06/09 (56817 MJD). 
We took one week integrations and selected the two energy ranges 100 MeV - 300 GeV and 300 MeV - 1 GeV.  
The data were searched again with LS and DC-DFT  (Table \ref{period} and Figure \ref{pow1}). 
It is apparent that the periodicity peaks at $T_{2} \sim$ 630-640 days,  which is consistent with twice 
the optical/NIR  period $T_{1}$. 
Note, however, that  the folded light curves cover a time interval of only $\sim$ 3 $\cdot\  T_{2}$, so this must be considered a tentative result at this time.
 Moreover the present data do not 
allow to comment about a possible phase shift of optical and $\gamma$-ray bands

\section{Discussion}

The simple fact that we recover the T$_{1} \sim$ 317 days period of \cite{Zhang2014} on independent data, 
though with an overlap of the observing time interval,  makes  their claim of this quasi-periodicity more robust.
The discovery of a possible quasi-periodicity at T$_{2} \sim$  2  T$_{1}$ in $\gamma$-rays, is intriguing, and per se an  
indirect confirmation of T$_{1}$. 
At this stage, because of the  limited number of monitored periods, it is impossible  to discriminate between
 a quasi-periodicity and a real period. 
Note that \textit{Fermi}  satellite follows this source  with no interaction with Sun constraints so that  the  light
 curve  is not annually biased. For the optical-NIR data,  T$_{1} \sim$ 317  days  can be distinguished from  1 yr period, 
 given the length of our  data-train and  this distinction is even stronger for the data   collected by
   \cite{Zhang2014}.  In fact in our Fourier spectra (Figure \ref{pow1}, top and central panels on the left) the power associated to a 1 yr periodicity is  significantly  lower  
    than that at $T_{1}$.
   In principle, it  is also possible that T$_{1}$ is a beat frequency for instance with a 1 yr  period. 
  In this case the ÒtrueÓ period could be $\sim$ 168 days (0.46 yr)   or  $\sim$ 2246 days ( 6.15 yr). 
   The former period does not appear in our power spectra, while the latter is comparable 
   with the duration of our optical observations.
     Further information on the quasi-periodicity may come from unpublished archived observations, 
     which we expect to be numerous since the source is a rather common target.
     Additional \textit{Fermi} data will show whether our tentative quasi-periodicity in  $\gamma$-rays holds 
     up long enough to be fully convincing.  Of interest would also be a detailed study of the X-ray light curves beyond that performed by \cite{Lachowicz2009}.
  This will be a delicate business. In fact because of the nature of most X-ray observations, 
 a combination of  data from numerous different instruments is required.  
 
  One can assume a mass of the black hole in PKS 2155-304 $\sim 10^{9}\  M_{\odot}$, as proposed by
\cite{Falomo1991} and 
\cite{Kotilainen1998} on the basis of the absolute magnitude of the host galaxy.  
If $T_{1}$  ($T_{2}$)  were a real periodicity one could consider  a Keplerian frame which yields  an 
orbital radius of $1.3 \cdot 10^{16}$ cm ($2 \cdot 10^{16}$ cm),  $\sim$ 40 (70) Schwarzschild radii.
The tidal effects on a star orbiting the supermassive black hole at such radius 
would be very relevant, a subject which deserves  further investigations.

\acknowledgments

We thank the referee for her/his prompt and competent report which enabled us to improve this letter.

We acknowledge the  support of the Italian Ministry of Education  (grant PRIN-MIUR 2009, 2010, 2011).
This work has also been supported by ASI grant I/004/11/0.

\end{document}